\documentclass[iop,apj]{emulateapj_tt}
\pdfoutput=1
\usepackage[breaklinks, colorlinks,
  urlcolor=blue, citecolor=blue, linkcolor=blue]{hyperref}
\usepackage{graphicx}
\usepackage{multirow}
\usepackage{color}
\usepackage{marginnote}
\reversemarginpar

\usepackage{fancyhdr}
\def \solar{_\odot}
\def \pow10#1{\times 10^{#1}}
\newcommand\blfootnote[1]{%
  \begingroup
  \renewcommand\thefootnote{}\footnote{#1}%
  \addtocounter{footnote}{-1}%
  \endgroup
}

\tighten

\begin{document}

\submitted{Submitted to ApJ Letters}

\title{Cookie-cutter halos: the remarkable constancy of the stellar
  mass function of satellite galaxies at $0.2<\lowercase{z}<1.2$}

\author{Tomer Tal\hyperlink{afflink}{$^{1}$}$^,$\hyperlink{afflink}{$^{6}$}}
\author{Ryan F. Quadri\hyperlink{afflink}{$^{2}$}}
\author{Adam Muzzin\hyperlink{afflink}{$^3$}}
\author{Danilo Marchesini\hyperlink{afflink}{$^4$}}
\author{Mauro Stefanon\hyperlink{afflink}{$^5$}}

\affiliation{\hypertarget{afflink}{$^1$ UCO/Lick Observatory, University of 
    California, Santa Cruz, CA 95064, USA;
    \href{mailto:tomer.tal@yale.edu}{tal@ucolick.org}}\\
  \hypertarget{afflink}{$^2$Department of Physics and Astronomy,
    Texas A\&M University, College Station, TX 77843, USA}\\
  \hypertarget{afflink}{$^3$ Leiden Observatory, Leiden University, 
    NL-2300 RA Leiden, The Netherlands}\\
  \hypertarget{afflink}{$^4$Department of Physics and Astronomy, Tufts 
    University, Medford, MA 02155, USA}\\
  \hypertarget{afflink}{$^5$Deptartment of Physics and Astronomy, 
    University of Missouri, Columbia MO, 65211, USA}}

\begin{abstract}
  We present an observational study of the stellar mass function of satellite
  galaxies around central galaxies at $0.2<z<1.2$.
  Using statistical background subtraction of contaminating sources we
  derive satellite stellar mass distributions in four bins of central 
  galaxy mass in three redshift ranges.
  Our results show that the stellar mass function of satellite galaxies 
  increases with central galaxy mass, and that the distribution of satellite
  masses at fixed central mass is at most weakly dependent on redshift.
  We conclude that the average mass distribution of galaxies in 
  groups is remarkably universal even out to $z=1.2$ and that it can be uniquely
  characterized by the group central galaxy mass.
  This further suggests that as central galaxies grow in stellar mass, they do
  so in tandem with the mass growth of their satellites.
  Finally, we classify all galaxies as either star forming or quiescent,
  and derive the mass functions of each subpopulation separately.
  We find that the mass distribution of both star forming and quiescent 
  satellites show minimal redshift dependence at fixed central mass.
  However, while the fraction of quiescent satellite galaxies increases 
  rapidly with increasing central galaxy mass, that of star forming satellites 
  decreases.\\

\end{abstract}



\section{Introduction}
\label{intro}
 \blfootnote{\hypertarget{afflink}{
     \hspace{-0.5cm}\noindent\rule{5.5cm}{0.4pt}\vspace{0.05cm}\\
     $^{6}$ NSF Astronomy and Astrophysics Postdoctoral Fellow}}
 Galaxy mass distributions in the universe have long been regarded as 
 an important characteristic of galaxy populations, and as a key tracer 
 of their evolution over time.
 Many of the physical processes that govern the evolution of galaxies directly 
 affect their masses, and as a consequence, also their global number density.
 Galaxy mergers, star formation and feedback mechanisms all leave their 
 imprint on the shape and normalization of the galaxy stellar mass function.
 Therefore, tremendous effort has been devoted to observational studies of 
 the global distribution of galaxy masses, pushing this measurement to higher 
 redshifts and to lower mass limits
 (e.g., Bell et al. \citeyear{bell_optical_2003}; 
 Bundy et al. \citeyear{bundy_mass_2006};
 Marchesini et al. \citeyear{marchesini_evolution_2009};
 Ilbert et al. \citeyear{ilbert_galaxy_2010};
 Muzzin et al. \citeyear{muzzin_evolution_2013};
 Tomczak et al. \citeyear{tomczak_galaxy_2014}).
 Similarly, the success of many numerical investigations of galaxy evolution is 
 determined in large part by their ability to reproduce the observed galaxy 
 stellar mass function 
 (e.g., Croton et al. \citeyear{croton_many_2006}; 
 Guo et al. \citeyear{guo_dwarf_2011}).

 Galaxies that reside in groups and clusters are subject to environmental 
 processes that affect their star formation rates and masses.
 Consequently, the distribution of galaxy masses in such halos is 
 somewhat different from the overall stellar mass function. 
 It is characterized by a gap between the mass (or luminosity) of central 
 galaxies and their satellite galaxies and it exhibits different fractions of
 star forming and quiescent galaxies in different environments
 (e.g., Jones et al. \citeyear{jones_nature_2003};
 Milosavljevi\'{c} et al. \citeyear{milosavljevic_cluster_2006};
 van den Bosch et al. \citeyear{van_den_bosch_towards_2007};
 Bolzonella et al. \citeyear{bolzonella_tracking_2010};
 Kova\v{c} et al. \citeyear{kovac_10k_2010};
 Vulcani et al. \citeyear{vulcani_galaxy_2013};
 Knobel et al. \citeyear{knobel_colors_2013};
 Deason et al. \citeyear{deason_stellar_2013};
 Van der Burg \citeyear{van_der_burg_environmental_2013}).
 In a series of papers, Yang et al. analyzed group membership catalogs using 
 extensive photometric and spectroscopic data from the Sloan Digital Sky
 Survey, to analyze the mass distribution of galaxies in groups at $z\sim0$.
 They found that the distribution of satellite galaxy masses can be reliably 
 predicted from the halo mass of a given group, or alternatively, from the 
 stellar mass of its central galaxy (the conditional stellar mass function; 
 Yang et al. \citeyear{yang_galaxy_2007}, \citeyear{yang_galaxy_2008}, 
 \citeyear{yang_galaxy_2009}).

 In this letter we follow an alternative approach to study the stellar mass 
 function of galaxies in groups in a large range of satellite and central 
 masses, and over a significant redshift range.
 Instead of assigning membership of individual galaxies to specific halos, we
 measure the average masses of galaxies in these environments using
 statistical background subtraction of contaminating sources.
 This technique has been shown to be effective in studies of satellite galaxies
 in general (e.g., Masjedi et al. \citeyear{masjedi_very_2006},
 \citeyear{masjedi_growth_2008};
 Tal et al. \citeyear{tal_observations_2012}, \citeyear{tal_galaxy_2013};
 Budzynski et al. \citeyear{budzynski_radial_2012};
 Nierenberg et al. \citeyear{nierenberg_luminous_2012}),
 and of the stellar mass function of satellites in
 particular (e.g., Tal \& van Dokkum \citeyear{tal_faint_2011};
 Tal et al. \citeyear{tal_mass_2012};
 Wang \& White \citeyear{wang_satellite_2012}).
 One of the main advantages to this approach is that it allows us to rely on 
 photometric surveys and thus probe the intermediate redshift universe using 
 mass limited samples.

 Throughout the letter we adopt the following cosmological parameters:
 $\Omega_m=0.3$, $\Omega_{\Lambda}=0.7$ and $H_0=70$ km s$^{-1}$
 Mpc$^{-1}$. 

 \begin{figure*}
   \includegraphics[width=1.0\textwidth]{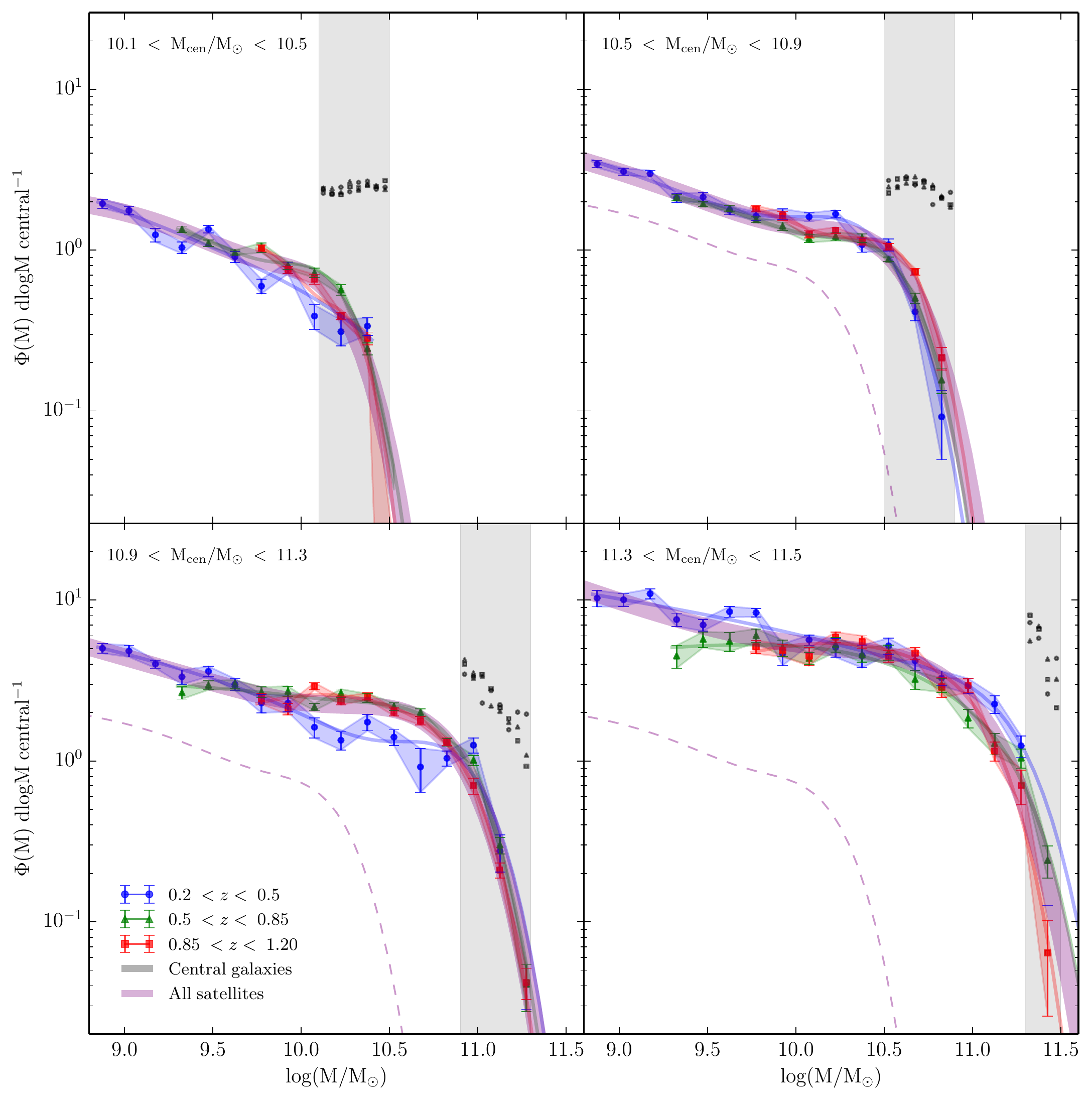}
   \caption{Redshift dependence of the satellite stellar mass function 
     at $0.2<z<1.2$ in four bins of fixed central galaxy mass.
     Black points and gray regions show the range of central galaxy masses 
     in each panel.
     Blue, green and red points represent satellite masses at different 
     redshifts, as well as their corresponding statistical uncertainties.
     Blue, green and red lines are double-Schechter fits to the data at each 
     redshift and thick purple lines are fits to the entire satellite population
     in each panel.
     The stellar mass functions of satellite galaxies at a given central 
     galaxy mass are consistent with one another across the studied redshift 
     range over essentially the entire spectrum of analyzed satellite masses.
     In contrast, the distribution of satellite masses varies significantly 
     with the mass of their central galaxy, in agreement with the conditional
     stellar mass function model (Yang et al. \citeyear{yang_galaxy_2009}).
     This weak dependence on redshift, and strong dependence on central mass
     imply that the average mass distribution of satellite galaxies can be 
     reliably predicted solely from the mass of their central at $0.2<z<1.2$.
   }
   \label{fig:bgall}
 \end{figure*}
 
 \begin{figure*}
   \includegraphics[width=1.0\textwidth]{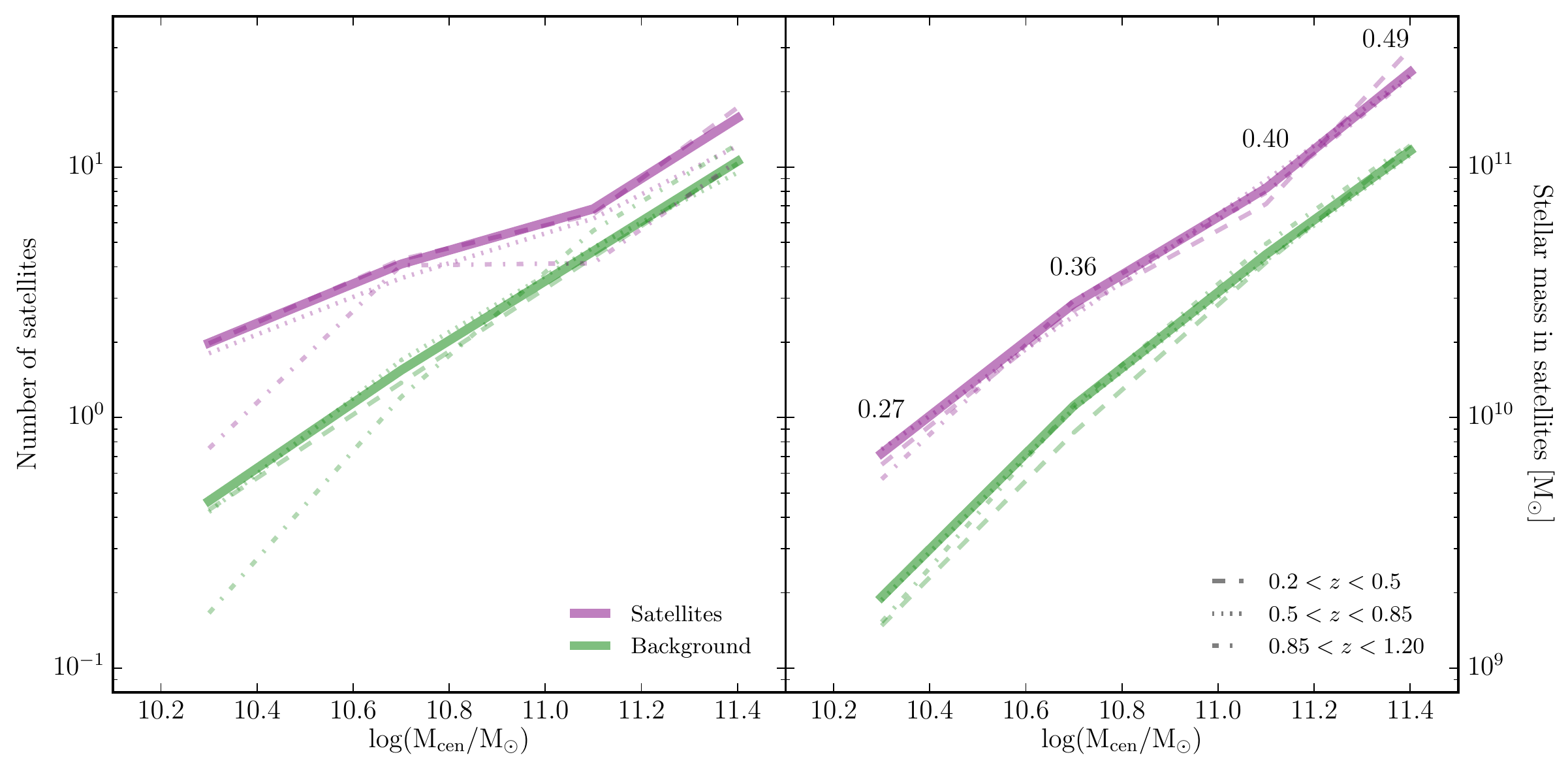}
   \caption{Number of satellites and their total stellar mass as a function
     of central galaxy mass.
     Solid purple lines represent the number of satellites (left panel) 
     and the total stellar mass enclosed in them (right panel) as 
     estimated by integrating double-Schechter fits to the satellite mass 
     functions at each central mass range (purple line in Figure 
     \ref{fig:bgall}).
     Dashed, dotted and dot-dashed lines show this calculation at each
     redshift separately.
     Green lines depict the same measurements for contaminating background 
     and foreground sources (as extracted from randomly positioned apertures).
     Broad spectral coverage of photometric data and consequent excellent 
     photometric redshifts in the UltraVISTA catalog allow us to keep a 
     relatively low contaminating source fraction ($\sim20\%-40\%$).
     Also noted in the figure are the average fractional contributions of 
     satellite mass to the overall mass budget in their halos.
     While satellite galaxies account for only $\sim27\%$ of the stellar
     mass in groups around low mass centrals, satellites around massive
     centrals contain as much stellar mass as the central galaxy itself.
   }
   \label{fig:fracbg}
 \end{figure*}
 
\section{Photometric catalog and Sample Selection}
\label{sec:data}
 Galaxies for this study were selected from the public photometric 
 catalog\footnote{\url{www.strw.leidenuniv.nl/galaxyevolution/ULTRAVISTA}}
 of Muzzin et al. (\citeyear{muzzin_public_2013}), based on the first data 
 release of UltraVISTA, an ongoing ultra deep near-infrared survey with the 
 European Southern Observatory VISTA survey telescope (McCracken et al. 
 \citeyear{mccracken_ultravista:_2012}).
 The catalog covers an area of 1.62 deg$^2$ in the COSMOS field, 
 includes photometry in 30 bands, and provides excellent photometric 
 redshifts ($\sigma_z/(1+z)=0.013$ with catastrophic outlier fraction of 
 1.56\%).
 Galaxy stellar masses are calculated assuming Kroupa 
 (\citeyear{kroupa_variation_2001}) initial mass function, and are estimated 
 to be 95\% complete down to a stellar mass limit of $5\pow10{9}M\solar$ at 
 $z<1.2$ (Muzzin et al. \citeyear{muzzin_evolution_2013}).
 
 \subsection{Central Galaxy Identification}
  We identified central galaxy candidates from the UltraVISTA catalog in
  three stellar mass bins of width 0.4 dex, in addition to one bin of 
  width 0.2 dex at the steep high mass end of the galaxy mass function.
  We then divided galaxies from the four mass bins into three redshift 
  intervals, spanning a total range of $0.2<z<1.2$.
  Galaxies were considered to be central if no other, more massive, galaxies
  could be found within two projected virial radii.
  Virial radius estimates at a given stellar mass and redshift were determined 
  using the semi-analytic model of Guo et al. (\citeyear{guo_dwarf_2011}).

 \subsection{Statistical Measurement of Satellite Galaxies}
  In order to study the mass distribution of satellite galaxies around the 
  selected centrals we followed the statistical background subtraction 
  procedure described by Tal et al. (\citeyear{tal_mass_2012}).
  We utilized photometric redshift measurements from the UltraVISTA catalog to 
  identify all galaxies within two projected virial radii from each central 
  that are separated from it by no more than $dz = 0.05$.
  This choice of aperture size was aimed at making a rather inclusive 
  selection of possible satellite galaxies.
  While some fraction of these sources are indeed physically associated with 
  the studied groups, other sources are potentially misidentified and instead 
  lie in the background or the foreground of the halo.
  We account for this contribution of contaminating sources by repeating the 
  same analysis in randomly positioned apertures in the field.

  For each central galaxy we selected a position at random from the area 
  allowed by the catalog coverage mask (as described in Muzzin 
  et al. \citeyear{muzzin_public_2013}).
  At this random position, we identified all galaxies that would be regarded as 
  satellite candidates according to virial radius estimates and redshift of
  the corresponding central.
  This procedure was repeated 20 times for each central galaxy.
 
 \begin{figure*}
   \includegraphics[width=1.0\textwidth]{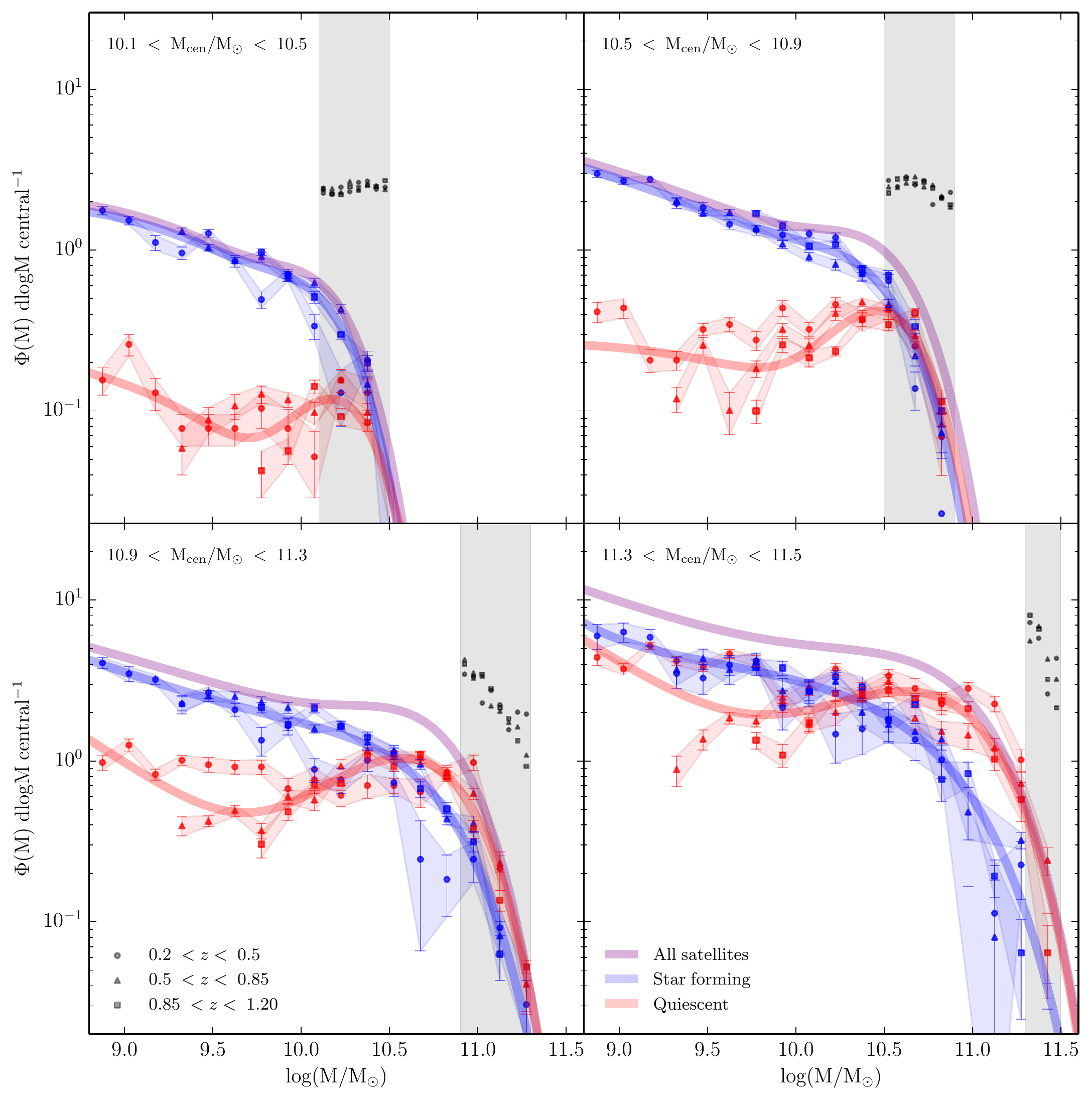}
   \caption{Stellar mass function of star forming and quiescent satellite 
     galaxies.
     Black points and gray regions show the range of central galaxy masses in
     each panel.
     Blue and red points represent the mass distribution of star forming and
     quiescent galaxies, respectively.
     Different marker symbols denote satellite mass functions at 
     different redshifts.
     Blue and red lines are double-Schechter fits to the star forming and 
     quiescent satellite data and the think purple line is a double-Schechter
     fit to the entire satellite population in each panel (same as in Figure
     \ref{fig:bgall}).
     The stellar mass functions of star forming galaxies at different redshifts
     are consistent with one another in the range $0.2<z<1.2$, while the mass 
     distributions of quiescent satellites are essentially constant with 
     redshift at $M_{\mathrm{sat}}/M_{\odot}\gtrsim10$.
     The relative contribution of each subpopulation to the overall
     satellite mass distribution strongly varies with central mass.
   }
   \label{fig:sfqall}
 \end{figure*}

\section{Satellite Galaxy Stellar Mass Functions}
\label{sec:mfunc}
 The stellar mass functions of satellite galaxies in the studied groups are
 revealed after removing the average contribution of contaminating 
 background and foreground sources.
 We derived the mass distribution of all sources that were found within 
 the central galaxy centered apertures at each redshift in each of 
 the four stellar mass bins.
 Similarly, we measured the mass functions of sources in randomly
 positioned apertures, thus quantifying the contribution from galaxies at the 
 same mass and redshift that are not associated with the selected centrals.
 The difference between the average mass functions in apertures 
 centered around centrals and in apertures centered around randomly selected 
 positions is taken to be the mass distribution of satellite galaxies that are 
 physically associated with the targeted groups.
 Error estimates in our satellite mass functions include statistical 
 uncertainties as were calculated from the 20 random apertures per
 central galaxy.

 \begin{figure*}
   \includegraphics[width=1.0\textwidth]{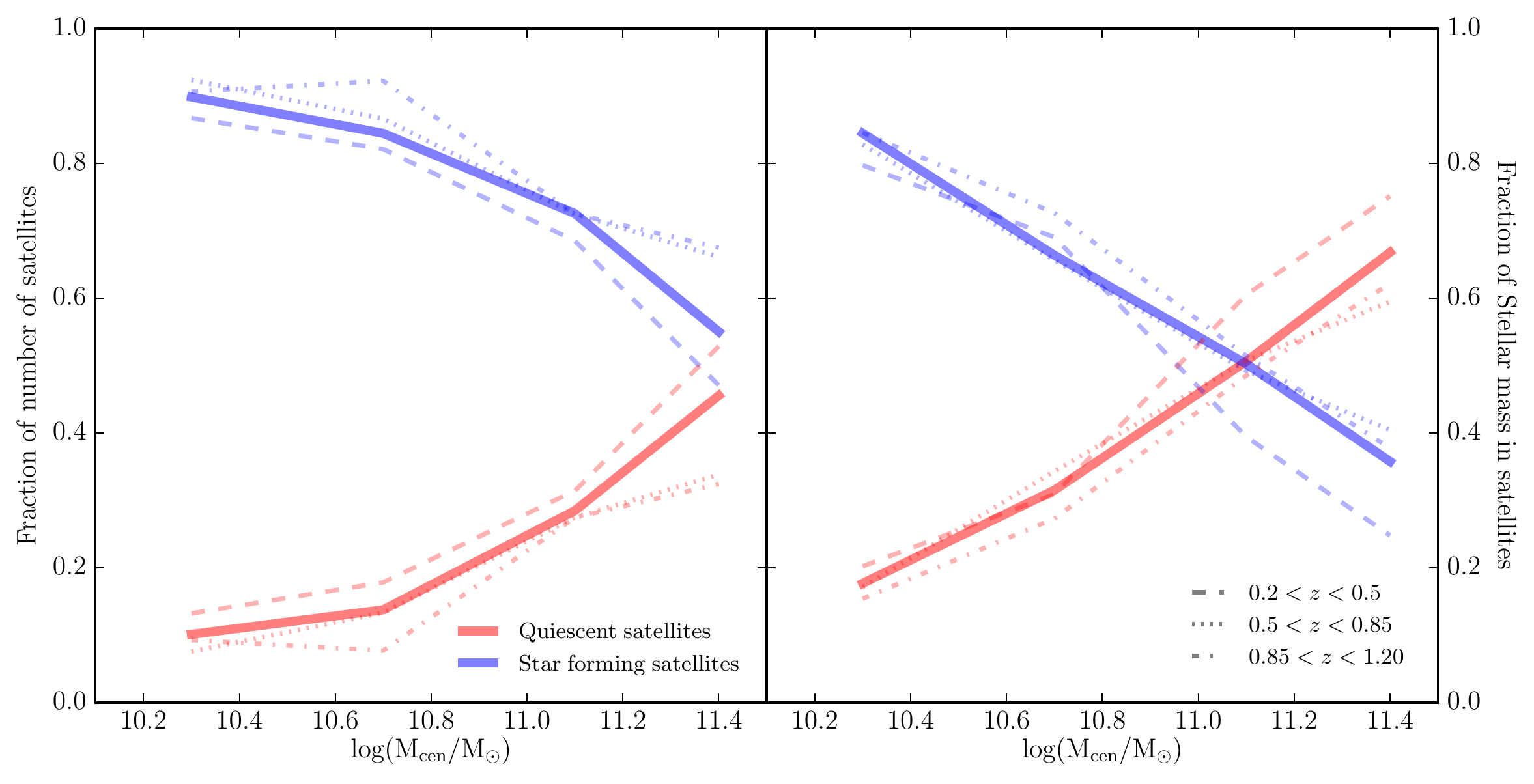}
   \caption{Relative contribution of star forming and quiescent satellite 
     galaxies.
     Blue and red lines show the fractions of satellite
     counts (left panel) and stellar mass (right panel) of star forming and
     quiescent satellite galaxies, respectively.
     The number of quiescent satellites nearly reaches that of star forming 
     satellites at the most massive end of the analyzed central mass range
     ($M_{\mathrm{cen}}/M_{\odot}\sim2.5\pow10{11}$).
     However, since quiescent satellites preferentially occupy the massive 
     end of the satellite mass function (as can be seen in Figure 
     \ref{fig:sfqall}), the total stellar mass enclosed in 
     them dominates over that of star forming satellites even in lower 
     mass halos (with $M_{\mathrm{cen}}/M_{\odot}\sim1.2\pow10{11}$).
   }
   \label{fig:fracbg2}
 \end{figure*}
 
 The results are shown in Figure \ref{fig:bgall}.
 Each panel shows the stellar mass function of satellite galaxies in three
 redshift intervals over a fixed range of central stellar mass.
 Completeness limits in stellar mass estimates define the plotted mass
 range at each redshift.
 The mass distributions of central galaxies are marked in each panel 
 as black points and a gray region.
 Solid blue, green and red lines are model fits to each 
 combination of central mass and redshift range, relying on the sum of two 
 Schechter (\citeyear{schechter_analytic_1976}) functions as the underlying 
 model.
 The thick purple line is a similar double-Schechter model fit to the entire
 satellite galaxy population in each panel.
 For comparison, the average mass function from the 
 $10.1 < \log(M_{\mathrm{cen}}/M_\odot) < 10.5$ panel is shown in the other 
 panels as a dashed purple line.
 Finally, the error bars in Figure \ref{fig:bgall} mark statistical 
 uncertainties as calculated from the standard deviation of the 20 random 
 mass functions per central galaxy.
 We note that the choice of double-Schechter model fitting was motivated by 
 results from recent studies, who found that such fits better 
 describe galaxy stellar mass functions than a single Schechter 
 function (e.g., Baldry et al. \citeyear{baldry_galaxy_2008}; 
 Li \& White \citeyear{li_distribution_2009};
 Muzzin et al. \citeyear{muzzin_evolution_2013};
 Tomczak et al. \citeyear{tomczak_galaxy_2014}).
 Best-fit model parameters are given in Table \ref{tab:fits}.

 \subsection{Lack of Redshift Dependence in the Distribution of 
   Satellite Masses at fixed central mass}
  The stellar mass functions of satellite galaxies at any of the four central
  galaxy mass ranges show no sign of significant dependence on redshift.
  As can be seen in Figure \ref{fig:bgall}, satellite mass distributions
  at different redshifts are consistent with one another within
  statistical uncertainties.
  By contrast, the overall distribution of satellite masses is strongly 
  dependent on the mass of their central galaxy (and by proxy also on the 
  group halo mass).
  This is evident from comparison of the satellite mass function at each 
  selection range of $M_{\mathrm{cen}}$ with that of the lowest central mass 
  bin (dashed purple curve in Figure \ref{fig:bgall}).
  The difference between the curves in each of the panels clearly shows that 
  while the slope of the low mass end of the satellite mass function stays 
  relatively unchanged, the overall normalization grows quickly with 
  increasing central mass.

  The strong dependence of the stellar mass function of satellite galaxies 
  on the mass of their central galaxy at $z=0$ has been demonstrated by 
  Yang et al. (\citeyear{yang_galaxy_2009}).
  Here we show that this dependence holds to at least $z=1.2$.
  Remarkably, this suggests that the mass distribution of satellite galaxies
  can essentially be characterized by a single parameter (either central 
  galaxy mass or its associated halo mass) over more than 60\% of the age of 
  the universe.
  The simplicity of this observed relation is accentuated by the complexity
  of processes that determine the distribution of satellite masses in 
  individual groups.
  Strikingly, the combined effect of such processes apparently does not 
  depend on redshift in the range $0.2<z<1.2$.


 \subsection{Contribution of Satellites to the Group Stellar Mass Budget}
  Figure \ref{fig:fracbg} shows the overall growth in the number and mass of 
  satellite galaxies with increasing central galaxy mass (purple lines).
  The total number density (left panel) and stellar mass density (right panel) 
  of satellite galaxies were derived by integrating each best-fit model from
  Figure \ref{fig:bgall} over the satellite mass range 
  $8.7 < \log(M/M_\odot) < 12.0$.
  For comparison, we also show the total number and mass of contaminating 
  sources as derived from randomly positioned apertures (green lines).
  The number and mass of satellite galaxies grow dramatically over the studied
  central mass range.
  Around the lowest mass centrals we find an average number of 2 satellites
  which make up roughly 27\% of the total stellar mass in the group.
  Around the most massive centrals this number increases to nearly 16 
  satellites, and the total stellar mass that is enclosed in them roughly
  equals the stellar mass of their centrals.
  As groups undergo hierarchical growth at $z<1$ (e.g., Williams et al. 
  \citeyear{williams_direct_2012}), they move up the track in Figure 
  \ref{fig:bgall}.
  This result suggests that the stellar mass in central and satellite galaxies
  must grow in tandem, and that the total mass in satellites grows at a faster
  rate than that of centrals.

  \begin{table*}[t]
    \caption{Double-Schechter best fit parameters}
    \centering
    \begin{tabular}{ c c c c c c }
      \hline\hline{\vspace{-8px}}\\

      \multicolumn{6}{c}{All galaxies} {\vspace{2px}}\\
      \hline
      {\vspace{2px}} Central mass & 
      $\log(M^*/M_\odot)$ & 
      $\alpha_1$ &
      $\log(\Phi^*_1)$ & 
      $\alpha_2$ & 
      $\log(\Phi^*_2)$ \\
      \hline  {\vspace{-7px}}\\

      \hspace{4mm} $10.1<M_{\mathrm{cen}}<10.5$ \hspace{4mm} & 
      \hspace{4mm} $9.61\pm0.01$ \hspace{4mm} & 
      \hspace{4mm} $1.55\pm0.52$ \hspace{4mm} & 
      \hspace{4mm} $0.29\pm0.04$ \hspace{4mm} & 
      \hspace{4mm} $-1.05\pm0.05$ \hspace{4mm} & 
      \hspace{4mm} $0.87\pm0.08$ \hspace{4mm} \\

      $10.5<M_{\mathrm{cen}}<10.9$ & $10.09\pm0.01$ & $0.83\pm0.13$ & 
      $0.91\pm0.04$ & $-1.28\pm0.01$ & $0.71\pm0.03$\\

      $10.9<M_{\mathrm{cen}}<11.3$ & $10.41\pm0.01$ & $-1.29\pm0.01$ & 
      $0.77\pm0.07$ & $0.42\pm0.10$ & $1.81\pm0.10$\\

      $11.3<M_{\mathrm{cen}}<11.5$ & $10.73\pm0.01$ & $-1.40\pm0.06$ & 
      $0.85\pm0.76$ & $-0.31\pm0.21$ & $3.67\pm0.47$\\
      \hline{\vspace{-8px}}\\

      \multicolumn{6}{c}{Star forming} {\vspace{2px}}\\
      \hline

      {\vspace{2px}}Central mass & $\log(M^*/M_\odot)$ & $\alpha_1$ & 
      $\log(\Phi^*_1)$ & $\alpha_2$ & $\log(\Phi^*_2)$ \\
      \hline {\vspace{-7px}}\\

      $10.1<M_{\mathrm{cen}}<10.5$ & $9.60\pm0.01$ & $1.44\pm0.64$ & 
      $0.27\pm0.04$ & $-1.04\pm0.06$ & $0.81\pm0.08$\\

      $10.5<M_{\mathrm{cen}}<10.9$ & $10.07\pm0.01$ & $0.56\pm0.27$ & 
      $0.69\pm0.03$ & $-1.30\pm0.02$ & $0.62\pm0.05$\\

      $10.9<M_{\mathrm{cen}}<11.3$ & $10.53\pm0.02$ & $-0.46\pm0.67$ & 
      $0.85\pm0.13$ & $-1.42\pm0.12$ & $0.34\pm0.28$\\

      $11.3<M_{\mathrm{cen}}<11.5$ & $10.81\pm0.01$ & $-1.13\pm0.02$ & 
      $1.26\pm0.11$ & $-2.91\pm30.6$ & $0.01\pm0.01$\\
      \hline{\vspace{-8px}}\\

      \multicolumn{6}{c}{Quiescent} {\vspace{2px}}\\
      \hline

      {\vspace{2px}}Central mass & $\log(M^*/M_\odot)$ & $\alpha_1$ & 
      $\log(\Phi^*_1)$ & $\alpha_2$ & $\log(\Phi^*_2)$ \\
      \hline {\vspace{-7px}}\\

      $10.1<M_{\mathrm{cen}}<10.5$ & $9.63\pm0.01$ & $2.53\pm0.96$ & 
      $0.02\pm0.01$ & $-1.17\pm0.14$ & $0.06\pm0.01$\\

      $10.5<M_{\mathrm{cen}}<10.9$ & $10.00\pm0.01$ & $1.96\pm0.55$ & 
      $0.14\pm0.01$ & $-1.00\pm0.11$ & $0.12\pm0.01$\\

      $10.9<M_{\mathrm{cen}}<11.3$ & $10.46\pm0.01$ & $0.33\pm0.04$ & 
      $1.15\pm0.01$ & $-1.72\pm0.06$ & $0.04\pm0.01$\\

      $11.3<M_{\mathrm{cen}}<11.5$ & $10.71\pm0.01$ & $-1.71\pm0.14$ & 
      $0.11\pm0.03$ & $-0.12\pm0.14$ & $3.03\pm0.16$\\
      \hline

    \label{tab:fits}
    \end{tabular}
    \footnotetext[1]{The single Schechter model is defined as: 
      $\Phi(M) = \ln(10)\Phi^*\left[10^{(M-M^*)(1+\alpha)}\right]
      \times \exp\left[−10^{(M-M^*)}\right]$}
  \end{table*}

\section{Stellar Mass Functions of Quiescent and Star Forming Satellites}
\label{sec:sats}
 In this section we analyze the mass distributions of star forming and 
 quiescent satellites separately.
 %
 We do so by classifying each galaxy in our sample as either ``star forming'' 
 or ``quiescent'', following the method described by 
 Williams et al. (\citeyear{williams_detection_2009}) and later successfully 
 repeated in numerous studies 
 (e.g., Brammer et al. \citeyear{brammer_number_2011}; 
 Szomoru et al. \citeyear{szomoru_sizes_2012};
 Barro et al. \citeyear{barro_candels:_2013}; 
 Muzzin et al. \citeyear{muzzin_evolution_2013};
 Tal et al. \citeyear{tal_observations_2014}).
 Williams et al. (\citeyear{williams_detection_2009}) showed 
 that galaxies out to $z=2$ can be reliably identified as having a low
 star formation rate based on a set of rest frame $U-V$ vs. $V-J$ ($UVJ$) 
 color selection criteria.
 Here we adopt the $UVJ$ threshold values that were found by Muzzin et al. 
 (\citeyear{muzzin_public_2013}) using redshift and rest frame color estimates 
 from the same UltraVISTA survey catalog that we utilize in this study.

 Figure \ref{fig:sfqall} shows the resulting stellar mass functions of star
 forming and quiescent satellites as a function of central galaxy mass and 
 redshift.
 As in Figure \ref{fig:bgall}, black points mark central galaxy masses and 
 thick purple lines follow the best-fit double-Schechter model to the entire
 sample in a given panel.
 Blue and red points and error bars show the mass distributions and statistical
 uncertainties of star forming and quiescent galaxies, respectively.
 Thick blue and red lines depict the best double-Schechter model fits to all
 galaxies in their respective subpopulation.

 \subsection{Rise of the Massive Quiescent Satellites}
  Similarly to the overall satellite mass functions, the mass distributions
  of both star forming and quiescent satellites exhibit no more than a weak 
  dependence on redshift.
  The stellar mass functions of star forming satellite galaxies at any given
  central mass are consistent with one another over the entire analyzed
  redshift range (blue lines in Figure \ref{fig:sfqall}).
  The same appears to be true for quiescent satellites at high masses
  ($\log(M/M_\odot) \gtrsim 10$), although there is possible evidence for 
  growth with time at lower masses (with large uncertainties).
  We note that evidence for different evolution at the low and 
  high end of the mass spectrum of quiescent galaxy populations has also 
  been found in studies of the global stellar mass function 
  (e.g., Peng et al. \citeyear{peng_mass_2010};
  Moustakas et al. \citeyear{moustakas_primus:_2013};
  Muzzin et al. \citeyear{muzzin_evolution_2013};
  Tomczak et al. \citeyear{tomczak_galaxy_2014}).

  The main difference between the mass functions of star forming and quiescent
  galaxies is in their relative contribution to the overall satellite mass
  distribution as a function of central mass.
  While the fraction of star forming galaxies at all satellite mass bins 
  decreases with increasing central mass, the fraction of quiescent satellites
  rises sharply.
  Moreover, the relative number of galaxies from each subpopulation does
  not vary evenly across the satellite mass range.
  Quiescent satellites preferentially occupy the high mass end of the stellar 
  mass function (at $\log(M/M_\odot) \gtrsim 10$) and make up most
  of the massive satellite population around massive centrals.
  At the same time, star forming galaxies account for most of the low mass
  satellites in all analyzed central galaxy mass bins.

  This can be seen in Figure \ref{fig:fracbg2}, where we calculate the 
  fractional contribution of star forming and quiescent galaxies to the 
  overall number density (top left) and mass density (top right) of satellites 
  as a function of central mass.
  As before, solid blue and red lines represent star forming and quiescent 
  galaxy subpopulations over $0.2<z<1.2$, and dashed, dotted 
  and dot-dashed lines show similar measurements at individual redshift bins.
  While star forming satellites make up the majority of the satellite galaxy
  population around centrals in all mass bins, the total mass in quiescent
  galaxies dominates the mass budget around massive centrals, at 
  $\log(M/M_{\mathrm{cen}}) \gtrsim 11.1$.
  
\section{Summary}
 The stellar mass function of satellite galaxies is an important ingredient 
 in galaxy evolution studies, as its shape and normalization are strongly 
 affected by several key processes in group and cluster halos.
 In this letter we utilized statistical background subtraction to derive 
 satellite mass functions in three fixed stellar mass samples of central 
 galaxies in the redshift range $0.2<z<1.2$.

 We showed that the mass distribution of satellite galaxies is independent 
 of redshift for any given value of central galaxy mass.
 In addition, since the satellite mass function increases strongly with the 
 mass of the central, this suggests that as groups grow with time, they move 
 along a universal central-to-total stellar mass relation.

 In addition, we integrated the mass functions of all samples and showed that 
 on average, the number of satellites increases from roughly 2 around central
 galaxies at $\log(M_{\mathrm{cen}}/M_\star) \sim 10.3$ to more than 16 at
 $\log(M_{\mathrm{cen}}/M_\star) \sim 11.4$.
 Furthermore, the total stellar mass that is enclosed in satellite galaxies 
 increases from roughly 27\% of the total group mass at the low mass end to
 nearly 50\%, where the most massive centrals in this study contain as much 
 stellar mass as their satellites.

 Finally, we derived stellar mass functions for star forming
 and quiescent satellites independently, and found that both are at most 
 weakly dependent on redshift at fixed central mass.
 We showed that the relative contribution of each subpopulation to the overall
 number and mass of satellites varies with central mass and that 
 quiescent satellites preferentially occupy the high mass end of the mass
 spectrum.
 As a result, the fraction of quiescent satellites reaches $\sim46\%$
 around the most massive subset of analyzed centrals 
 ($M_{\mathrm{cen}}/M_\odot\sim11.4$), where more than 65\% of the total 
 satellite mass is locked in quiescent galaxies.
 
 This is the first time that such an analysis is performed at the stellar mass
 and redshift ranges that are presented here, and it demonstrates that despite
 the complexity and large range of processes that govern satellite galaxy 
 evolution, the resulting average satellite mass function follows a simple 
 relation with central mass out to $z\sim1.2$.


\begin{acknowledgements}
%
  We thank Kim-Vy Tran for engaging discussions which contributed to this work.

  TT is supported by an NSF Astronomy and Astrophysics Postdoctoral Fellowship 
  under award AST-1202667.

  This study is based in part on a K$_{s}$-selected catalog of the 
  COSMOS/UltraVISTA field from Muzzin et al. (\citeyear{muzzin_public_2013}).  
  The catalog contains PSF-matched photometry in 30 photometric bands covering 
  the wavelength range 0.15$\micron$ $\rightarrow$ 24$\micron$ and includes 
  the available $GALEX$ (Martin et al. \citeyear{martin_galaxy_2005}), 
  CFHT/Subaru (Capak et al. \citeyear{capak_first_2007}), 
  UltraVISTA (McCracken et al. \citeyear{mccracken_ultravista:_2012}), 
  S-COSMOS (Sanders et al. \citeyear{sanders_s-cosmos:_2007}), and 
  zCOSMOS (Lilly et al. \citeyear{lilly_zcosmos_2009}) datasets.
  The catalog was derived using data products from observations made with ESO 
  telescopes at the La Silla Paranal Observatory under ESO programme 
  ID 179.A-2005 and on data products produced by TERAPIX and the Cambridge 
  Astronomy Survey Unit on behalf of the UltraVISTA consortium.
\end{acknowledgements}

\bibliographystyle{yahapj}
\bibliography{ms}

\end{document}